\documentclass[twocolumn]{aastex6}

\usepackage{amsmath}
\usepackage{natbib}
\usepackage{filecontents}

\newcommand{\RNum}[1]{\uppercase\expandafter{\romannumeral #1\relax}}
\usepackage{romannum}
\begin{document}


\title{The orbit of the gamma-ray binary 1FGL J1018.6$-$5856}


\author{I.M. Monageng\altaffilmark{1,2}$^{\ast}$, V.A. McBride\altaffilmark{1,2,3}, L.J. Townsend\altaffilmark{2}, A.Y. Kniazev\altaffilmark{1,5,6,7}, S. Mohamed\altaffilmark{1,2,8} and M. B\"{o}ttcher\altaffilmark{4}}
\affil{1. South African Astronomical Observatory, P.O Box 9, Observatory, 7935, Cape Town, South Africa}
\affil{2. Department of Astronomy, University of Cape Town, Private Bag X3, Rondebosch 7701, South Africa}
\affil{3. Office of Astronomy for Development, IAU, Cape Town, South Africa}
\affil{4. Centre for Space Research, North-West University, Potchefstroom, 2531, South Africa}
\affil{5. Southern African Large Telescope Foundation, PO Box 9, 7935 Observatory, Cape Town, South Africa}
\affil{6. Sternberg Astronomical Institute, Lomonosov Moscow State University, Universitetskij Pr. 13, Moscow 119992, Russia}
\affil{7. Special Astrophysical Observatory of RAS, Nizhnij Arkhyz, Karachai-Circassia 369167, Russia}
\affil{8. South Africa National Institute for Theoretical Physics, Private Bag X1, Matieland, 7602, South Africa}

\begin{abstract}
Gamma-ray binaries are a small subclass of the high mass X-ray binary population which exhibit emission across the whole electromagnetic spectrum.  We present radial velocities of 1FGL J1018.6$-$5856 based on observations obtained with the Southern African Large Telescope (SALT). We combine our measurements with those published in the literature to get a broad phase coverage. The mass function obtained supports a neutron star compact object, although a black hole mass is possible for very low inclination angles. The improved phase coverage allows constraints to be placed on the orbital eccentricity ($e$ = 0.31  $\pm$  0.16), which agrees with estimates from high energy data. 
\end{abstract}




\section{Introduction} \label{sec:intro}

Gamma-ray binaries (GRBis) are a small subclass of high mass X-ray binaries (HMXBs), comprising only six sources: PSR B1259$-$63 \citep{1992MNRAS.255..401J}, LS 5039 \citep{1997A&A...323..853M}, LS I+61 303 \citep{1978IAUC.3164....1G}, HESS J0632+057 \citep{2007A&A...469L...1A}, 1FGL J1018.6$-$5856 \citep{2011HEAD...12.0307C} and LMC P3 \citep{2016ApJ...829..105C}. These systems comprise a pulsar or black hole (BH) and a massive O or Be star. GRBis differ from traditional HMXBs (Be X-ray and supergiant X-ray binaries) in that they show a peak in their spectral energy distribution (SED) above 1~MeV, as well as display multiwavelength emission across the whole electromagnetic spectrum. In all but one case (PSR B1259$-$63) the nature of the compact object is uncertain (it is either a neutron star or BH). Through radio pulse timing, a pulse period of 47.7\,ms was found in PSR B1259$-$63, confirming the nature of the compact object as a pulsar. Pulsations are not detected in any of the other gamma-ray binaries. Furthermore, the large uncertainty in the mass function results in a large uncertainty in the calculated mass of the compact object in the other systems, making it difficult to distinguish between a pulsar and a BH (a minimum compact object mass of 3\,M$_{\odot}$ would indicate a BH). \\
1FGL J1018.6$-$5856 (J1018), the subject of this paper, was discovered at GeV energies when a 16.6\,day modulation was detected from the first Fermi/LAT catalog \citep{2011HEAD...12.0307C}. Follow-up work at optical wavelengths taken with the 1.9\,m telescope at the South African Astronomical Observatory (SAAO) and the 2.5\,m telescope at Las Campanas Observatory (LCO) revealed Balmer, He~\textsc{\romannum{1}} and He~\textsc{\romannum{2}} absorption lines (similar to LS 5039), with the He~\textsc{\romannum{2}}/He~\textsc{\romannum{1}} ratio indicating an O6V((f)) spectral type \citep{2012Sci...335..189F}.\\
 Using \'Echelle spectra obtained with the CTIO 1.5\,m telescope, \cite{2015ApJ...812..178W} performed a radial velocity (RV) analysis of J1018. Using Gaussian fits to the H~\textsc{\romannum{1}}, He~\textsc{\romannum{1}} and He~\textsc{\romannum{2}} absorption lines, \cite{2015ApJ...812..178W} found a semi-amplitude ($K$) range of 12$-$40\,km/s. The maximum determined semi-amplitude (40\,km/s) from their work means that the compact object likely has a mass $>$2.2\,M$_{\odot}$, favouring a BH. The RV curve from \cite{2015ApJ...812..178W}, however, is hindered by a large scatter due to imprecision of the continuum calibration, as well as possible contamination by stellar wind line features. \cite{2015ApJ...813L..26S} (S15) performed a follow-up RV study of J1018 using medium resolution spectra obtained with the Goodman High-Throughput Spectrograph onboard the SOAR 4.1\,m telescope. They performed a cross-correlation of their spectra against the O6V star, HD 172275, in two wavelength regions with the H$\gamma$ line and He~\textsc{\romannum{2}} lines (4542\,\AA \textrm{ }and 4686\,\AA). The two RV plots show different systemic velocities, with the He~\textsc{\romannum{2}} RV measurements offset from the Balmer line measurements by +6~km/s. A similar behaviour from RV measurements made from the two line species has been reported in LS 5039, which is believed to be due to the contamination of wind for the He~\textsc{\romannum{2}} and H~\textsc{\romannum{1}} \citep{2005MNRAS.364..899C,2011MNRAS.411.1293S}. With a modest phase coverage, S15 performed a circular Keplerian fit to the RV measurements and found semi-amplitudes of 11.4 $\pm$ 1.5\,km/s and 12.2 $\pm$ 2.7\,km/s from H$\gamma$ and He~\textsc{\romannum{2}}, respectively. S15 found that for inclination angles between 26$^{\circ}$ and 64$^{\circ}$ a compact object mass of 1.4$-$2.5\,M$_{\odot}$ (canonical neutron star mass range) is obtained, while for lower inclination angles ($i \leq$ 16$^{\circ}$) a BH is favoured (M$_{X} \geq$ 5M\,$_{\odot}$). \\
 In this work, we present new \'Echelle optical spectroscopy of J1018 obtained with the High Resolution Spectrograph (HRS) on the Southern African Large Telescope (SALT). The broader phase coverage, combined with radial velocities from S15, allows for better constraints on the orbital parameters and hence mass estimate of the compact object.


\begin{figure}
	\centering
	\includegraphics[width = 0.35\textwidth, angle=270]{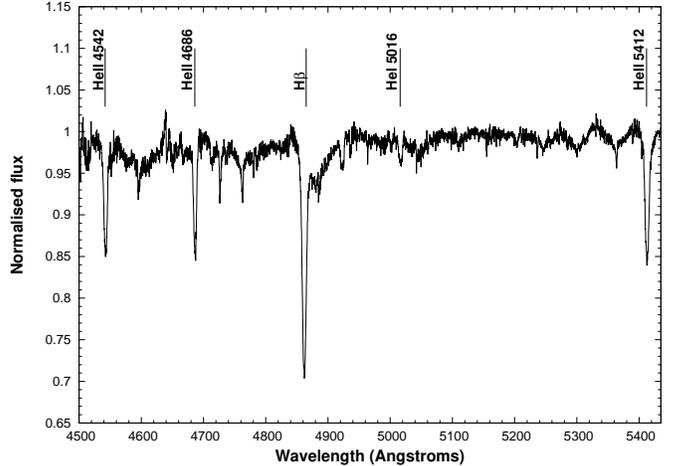}
	\caption{Normalized average blue spectrum for the wavelength range used of 1FGL J1018.6$-$5856 with the strongest lines used for cross-correlation labelled.}
	\label{spec}
\end{figure}

\section{Observations and reductions}
J1018 was observed 8 times with SALT \citep{2006IAUS..232....1B} using the HRS \citep{2010SPIE.7735E..4FB,2012SPIE.8446E..0AB,2014SPIE.9147E..6TC} between 06 December 2015 and 20 December 2015. Our observations were done in medium resolution (MR) mode ($\textrm{R}\sim40000$) with the blue and red arms, providing a total wavelength coverage of $\sim$3700$-$8900\,\AA. Only spectra with wavelength ranges 4500$-$5500\,\AA~and 5700$-$6750\,\AA~for the blue and red arms, respectively, were useable. All observations were performed with exposure times of 1400\,s. Regular calibration sets of ThAr arc lamps, bias and flats were taken throughout the duration of our programme. Primary reductions of the spectra (including overscan correction, bias subtraction and gain correction) were performed with the SALT science pipeline \citep{2015ascl.soft11005C}. The subsequent reduction steps which include background subtraction, removal of the blaze function, identification of the arc lines and merging of the orders for the object spectra were carried out using the \textsc{midas feros} \citep{1999ASPC..188..331S} and \textsc{echelle} \citep{1992ESOC...41..177B} packages (see \citealt{2016MNRAS.459.3068K} for a detailed description of the reduction steps). The spectra were normalised and corrected for the heliocenter using \textsc{iraf}\footnote{Image Reduction and Analysis Facility: iraf.noao.edu} tasks \textsc{rvcorrect} and \textsc{dopcor}. 

\section{Radial velocities}
\begin{figure}
	\centering
	\includegraphics[width = 0.5\textwidth]{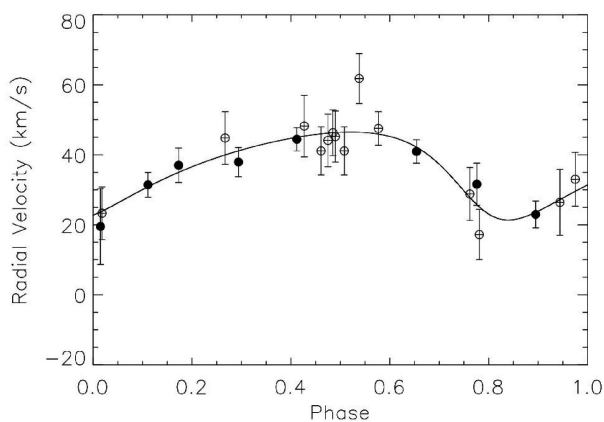}
	\caption{Best-fitting curve to the radial velocities of the He~\textsc{\romannum{2}} lines. The unfilled circles are from Strader et al. (2015) and the filled circles are from this work.}
	\label{RV_fit}
\end{figure}

For our analysis we use the blue spectra to measure radial velocities. Figure \ref{spec} shows the mean spectrum of J1018 with the strongest absorption lines of He~\textsc{\romannum{1}} (5016~\AA), He~\textsc{\romannum{2}} (4542, 4686 and 5411~\AA) and H$\beta$ present. For the wavelength range that we have access to, the mean spectrum is consistent with the low resolution spectrum of \cite{2012Sci...335..189F}. To obtain the radial velocities we performed a cross-correlation using the blue arm spectra. We used an iterative process described by \cite{2015MNRAS.448.1789M} and \cite{2003MNRAS.338..360F} to generate a high signal-noise ratio (SNR) zero velocity template. First, individual spectra were rectified and the continuum level subtracted. Each spectrum was then converted to a logarithmic wavelength scale and the spectra were grouped according to SNR (highest to lowest). The radial velocity (RV) shifts of each spectrum relative to the spectrum with the highest SNR were measured. Individual spectra were then shifted to the same rest wavelength as the high SNR template using the results of the first cross-correlation iteration, and a mean spectrum created by combining all the shifted spectra. This became our final template spectrum, which was then used to compute the RV shifts. These shifts (converted to velocities) are shown in Table~\ref{RV_table} and Fig.~\ref{RV_fit} (filled circles). \\

\begin{table}
\centering
 \caption{Radial velocities of 1FGL J1018.6$-$5856 from SALT.}
 \label{tab1}
 \begin{tabular}{l l}
  \hline
  BJD (days) & Velocity (km/s)  \\ \hline
2457362.5348	&	31.4  $\pm$  3.6 \\
2457363.5595	&	36.9  $\pm$  4.9 \\
2457365.5589	&	37.9  $\pm$  4.2 \\
2457367.5202	&	44.4  $\pm$  3.4 \\
2457371.5214	&	40.9  $\pm$  3.3 \\
2457373.5396	&	31.6  $\pm$  6.0 \\
2457375.5024	&	22.9  $\pm$ 3.9 \\
2457377.4990	&	19.5  $\pm$ 10.9 \\ \hline
\label{RV_table}
\end{tabular}
\end{table}

Different combinations of the line species were considered during the cross-correlation analysis, however no significant differences in the RV amplitudes were found. The final values of RV used are those from all the available lines in the wavelength range 4500$-$5500\,\AA. The measurements obtained from this work are combined with those from S15. S15 consider the two line species of He~\textsc{\romannum{2}} (4542 and 4686 \AA) and H$\beta$ separately. Previous RV studies involving O stars have considered the two line species separately \citep{2005MNRAS.364..899C, 2011MNRAS.411.1293S} in the case of LS 5039. The Balmer lines are more contaminated by the wind emission than the He~\textsc{\romannum{2}} lines, which originate from the stellar photosphere \citep{1996A&A...305..171P, 2011MNRAS.411.1293S, 2015ApJ...805...18W}, making He~\textsc{\romannum{2}} lines more reliable in the derivation of orbital parameters. For the reasons explained above we use the combined RV values from our work, using the full wavelength range of the blue spectra, and the He~\textsc{\romannum{2}} measurements from S15 to derive the final orbital solutions. A Keplerian model (Fig.~\ref{RV_fit}) was fitted to the combined measurements with the period fixed to that derived from X-ray studies of J1018 ($P = 16.544$ days, \citealt{2015ApJ...806..166A}). We used the \textsc{rvlin} software package provided by \cite{2009ApJS..182..205W} to perform the Keplerian fits and obtain the model parameters. The uncertainties in the best-fit parameters were obtained using the \textsc{boottran} bootstrapping routines described in \cite{2012ApJ...761...46W}. We note that if the period is left as a free parameter, a period of $P = 16.583 \pm 0.042$ days is derived, in good agreement with the X-ray period (as well as that derived from gamma-rays, $P = 16.549 \pm 0.007$ days; \citealt{2014HEAD...1412210C}). A summary of the orbital parameters is presented in Table ~\ref{params}.

\begin{table}
\centering
 \caption{Orbital elements of 1FGL J1018.6$-$5856}
 \label{tab1}
 \begin{tabular}{l l}
  \hline
  Parameter & Value \\ \hline
  $P_\mathrm{orb}$ (days) & 16.544 (fixed) \\ 
  $T_\mathrm{p}$ (JD)  & 2457256.0  $\pm$  1.2\\ 
  $e$ & 0.31  $\pm$  0.16 \\ 
  $\omega (^\circ)$  & 89  $\pm$  30 \\ 
  $K$ (km/s) & 12.3  $\pm$  1.9 \\ 
  $\gamma$ (km/s) & 35.5  $\pm$  1.3 \\ 
  RMS of fit (km/s) & 5.85  \\ \hline
  \label{params}
 \end{tabular}
\end{table}

\section{Mass of the compact object}
\begin{figure}
	\centering
	\includegraphics[width = 0.35\textwidth, angle=270]{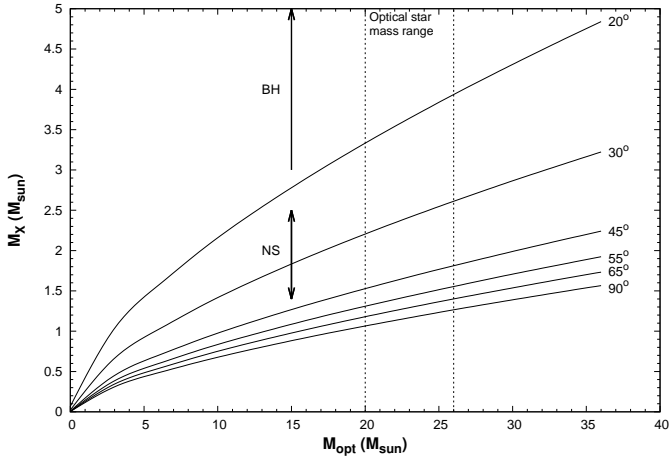}
	\caption{Mass constraints for the two systems in 1FGL J1018.6$-$5856 for different inclination angles. The two vertical lines show the mass range for the optical star.}
	\label{massfun}
\end{figure}

The derived orbital parameters allow us to use the mass function defined by

\begin{equation}
	f(M_{\textrm{x}}) = \frac{PK^3}{2\pi G}(1-e^2)^{3/2} = \frac{(M_{\textrm{x}} \sin i)^3}{(M_{\textrm{x}} + M_{\textrm{O}})^2}, 
\end{equation}

where $M_{\textrm{x}}$ and $M_{\textrm{O}}$ are the masses of the compact object and optical companion, respectively, and $i$ is the inclination angle of the orbit. For the remainder of the analysis we use an optical companion mass range for an O6V((f)) of 20$-$26.4 M$_{\odot}$ as used by \cite{2005MNRAS.364..899C}, \cite{2015ApJ...813L..26S} and \cite{2015ApJ...812..178W}. For the parameters listed in Table~\ref{params} a mass function of $f(M_{\textrm{x}}) = 0.0027 \pm 0.0013$ M$_{\odot}$ is obtained. From this, the mass of the compact object can be calculated for different inclination angles. Fig.~\ref{massfun} shows the mass-mass plot obtained from the orbital parameters derived from the RV fit. Referring to Fig.~\ref{massfun}, for a lower limit mass of the optical companion (20 M$_{\odot}$) and a canonical neutron star mass range of 1.4$-$2.5 M$_{\odot}$, this implies an orbital inclination angle range of $\sim$50$-$26$^{\circ}$. For the upper limit optical companion mass of 26.4 M$_{\odot}$, inclination angles between $\sim$70$^{\circ}$ and 32$^\circ$ are implied. The most massive NS known has a mass of 2.0~M$_\odot$ \citep{2013Sci...340..448A}, which corresponds to inclination angles of $i = 33^\circ$ and $i = 41^\circ$ for optical companion mass of 20~M$_\odot$ and 26.4~M$_\odot$, respectively. For the compact object to be a BH ($M_{\textrm{x}} \geq 3.0$ M$_{\odot}$), lower inclination angles of $i < 22^{\circ}$ and $i < 26^{\circ}$ for optical companion mass of 20 M$_{\odot}$ and 26.4 M$_{\odot}$, respectively, are implied. 
\section{Discussion}
\begin{figure}
	\centering
	\includegraphics[width = 0.45\textwidth]{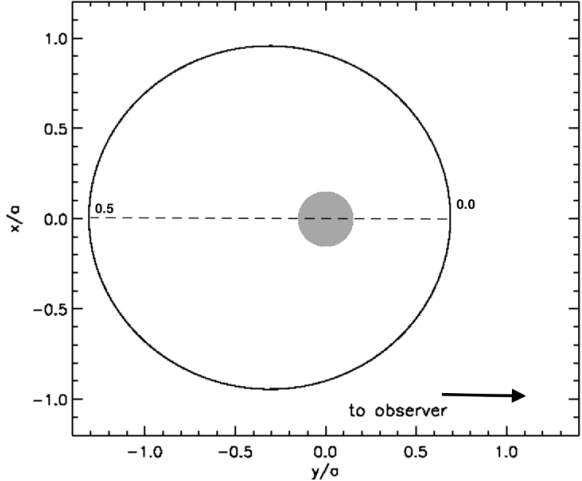}
	\caption{The orbit of the compact object around the optical companion (shaded circle) in J1018 as seen from above the orbital plane. Periastron and apastron phases are indicated in the plot. The coordinates are in units of the semimajor axis.}
	\label{geometry}
\end{figure}

The value of the mass function obtained from this study agrees with that found by \cite{2015ApJ...813L..26S} to within the errors. For the mass range of the companion star considered, we also find similar values for the allowed inclination angles for different masses of the compact object (Fig.~\ref{massfun}). We obtained a best-fit Keplerian model of eccentricity $e = 0.31  \pm  0.16$. The probability of this model, given our data, is $P(\chi^2 > 11.05)\sim$85~\% (a circular fit results in a probability $\sim$50\%). The analysis of Fermi GeV data revealed a relatively low modulation amplitude which, if the gamma-ray flux is due to anisotropic inverse Compton scattering of stellar photons by energetic electrons, implies low inclination and eccentricity \citep{2017arXiv170308080C}. \cite{2017ApJ...838..145A} model the X-ray and gamma-ray light curves and SED of J1018, where the fit of the model to the spike in the modulated X-ray light curve is explained by an orbital eccentricity of $e = 0.35$ and inclination angle of $i \sim 50^{\circ}$. These implications for the inclination angle from high energy analysis further support a neutron star as the compact object in J1018. Furthermore, the mass of the donor star could be significantly less than the estimates used, which is typical of donors in HMXBs to display different characteristics than those of a main-sequence equivalent (e.g. \citealt{2015MNRAS.447.2387C,2017MNRAS.464.4133R} and references therein). This would reduce the mass estimate of the compact object for a given mass function, making a NS even more likely. The wide phase coverage allows us to put constraints on the orbital eccentricity, which is also in agreement (to within errors) with estimates from high energy studies \citep{2017ApJ...838..145A}. Fig.~\ref{geometry} shows the orbit of the compact object around the optical companion as viewed from an inclination angle of $i = 0^{\circ}$. This was produced using mass and radius of 22.9~M$_\odot$ and 9.3~R$_\odot$, respectively, for the optical companion \citep{2005MNRAS.364..899C} and a compact object of mass 1.4~M$_{\odot}$. From Fig.~\ref{geometry}, the periastron passage of the compact object occurs close to inferior conjunction (INFC). The peak of the X-ray, GeV and TeV emission in J1018 occurs at similar phases \citep{2015A&A...577A.131H, 2015ApJ...806..166A}. If the gamma-ray emission is due to anisotropic inverse Compton scattering of stellar photons, then the GeV peak is expected to occur at phases when the compact object is behind the donor star (superior conjunction, SUPC), while TeV emission is expected to be maximum at INFC (since absorption is maximum at SUPC). The peculiar behaviour of the emission maxima resulting at similar phases is discussed by S15, where a proposed solution is the occurrence of INFC and periastron passage at a similar phase.
 
\begin{figure}
	\centering
	\includegraphics[width = 0.54\textwidth, angle=0]{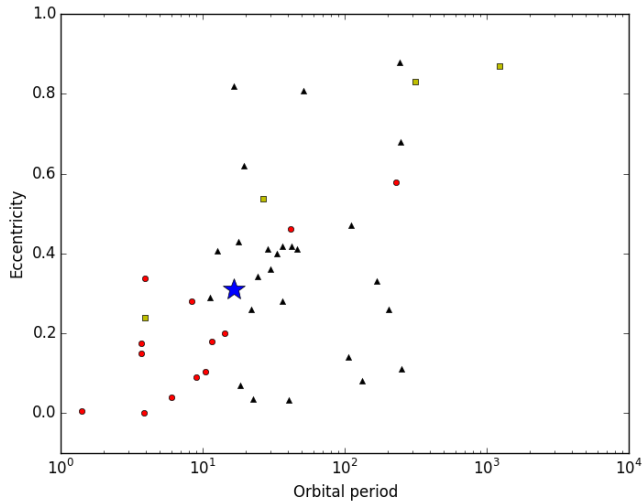}
	\caption{Updated orbital period against eccentricity for HMXBs from \cite{2011MNRAS.416.1556T}. The red circle, black triangle and yellow square symbols represent supergiant, Be X-ray and GRBi systems, respectively. The blue star represents J1018.}
	\label{ecc_per}
\end{figure}

Fig.~\ref{ecc_per} shows the updated orbital period vs. eccentricity plot from \cite{2011MNRAS.416.1556T} for HMXBs with J1018 included from the derived orbital eccentricity in this work. J1018 lies around the transition zone between supergiant and Be X-ray binary systems, and follows the correlation of the two quantities which was noted by \cite{2012MNRAS.421.1103C} for the GRBis  with known orbital parameters. It was speculated that the correlation between eccentricity and orbital period in GRBis is due to the small separation required for very high energy (TeV) emission to be triggered, which for long period systems requires large eccentricities. With the limited sample, GRBis display the same characteristics as accreting HMXBs, supporting the notion that GRBis may represent an earlier phase in the evolution of HMXBs.  

\section{Conclusion}
We observed J1018 with SALT and, combining our observations with S15, we obtain orbital phase coverage which allows us to constrain the orbital parameters. In particular, we obtained an eccentricity of $e = 0.31  \pm  0.16$ and a mass function $f(M_{\textrm{x}}) = 0.0027 \pm 0.0013$ M$_{\odot}$. The eccentricity obtained is in agreement with that implied from high energy studies of the source. For a range of values of the mass of the optical companion of $20-26.4$~M$_{\odot}$, the mass function obtained gives a mass for the compact object which favours a NS, with a BH possible for very low inclination angles ($i \lesssim 26^{\circ}$).

\section*{Acknowledgments}
We thank the anonymous referee for helpful comments that improved the paper. All spectral observations reported in this paper were obtained with the South African Large Large Telescope under the proposal code 2015-2-SCI-045 (PI: Monageng). We thank Brent Miszalski and Yuki Moritani for the useful discussions related to cross-correlation techniques. IM acknowledges funding from the UCT science faculty PhD fellowship. VM acknowledges support of the National Research Foundation of South Africa (Grant numbers 98969 and 93405 ). AK acknowledges the National Research Foundation of South Africa and the Russian Science Foundation (project no. 
14-50-00043). SM is grateful to the South African National Research Foundation (NRF) for a research grant. The work of MB is supported by the South African Department of Science and Technology and National Research Foundation through the South African Research Chairs Initiative.
\software{SALT science pipeline \citep{2015ascl.soft11005C}, Feros \citep{1999ASPC..188..331S}, echelle \citep{1992ESOC...41..177B}, IRAF \citep{1986SPIE..627..733T,1993ASPC...52..173T}, rvlin \citep{2009ApJS..182..205W}, boottran \citep{2012ApJ...761...46W}}
\bibliography{references}

\listofchanges

\end{document}